\documentclass[aps,prl,twocolumn,superscriptaddress,showpacs]{revtex4-1}

\usepackage{graphicx}
\usepackage{amsmath, amsfonts, amssymb, amsbsy}
\usepackage{MnSymbol}

\begin{document}

%\preprint{}

\title{High-pressure spin-shifts in the pseudogap regime of superconducting YBa$_{2}$Cu$_{4}$O$_{8}$ revealed by $^{17}$O nuclear magnetic resonance}


\author{Thomas Meissner}
%\email[]{meissner@physik.uni-leipzig.de}
\affiliation{Faculty of Physics and Earth Science, University of Leipzig, Germany}
\author{Swee K. Goh}
\affiliation{Department of Physics, Cavendish Laboratory, University of Cambridge, United Kingdom}
\author{J\"urgen Haase}
\affiliation{Faculty of Physics and Earth Science, University of Leipzig, Germany}
\author{Grant V. M. Williams}
\affiliation{The MacDiarmid Institute and Industrial Research Limited, New Zealand}
\author{Peter B. Littlewood}
\affiliation{Department of Physics, Cavendish Laboratory, University of Cambridge, United Kingdom}


\date{\today}

\begin{abstract}
A new NMR anvil cell design is used for measuring the influence of high pressure on the electronic properties of the high-temperature superconductor YBa$_2$Cu$_4$O$_8$ above the superconducting transition temperature $T_{\rm c}$. It is found that pressure increases the spin shift at all temperatures in such a way that the pseudo-gap feature has almost disappeared at 63~kbar. This change of the temperature dependent spin susceptibility can be explained by a pressure induced proportional decrease (factor of two) of a temperature dependent  component, and an increase (factor of 9) of a temperature independent component, contrary to the effects of increasing doping. The results demonstrate that one can use anvil cell NMR to investigate the tuning of the electronic properties of correlated electronic materials with pressure. 
\end{abstract}

\pacs{74.25.nj, 74.62.Fj, 74.72.Kf}


\keywords{Superconductivity, NMR, High Pressure, Pseudo-Gap}

\maketitle

For the investigation of the rich properties of correlated electronic materials not only, e.g., temperature or magnetic field, but also pressure is a very useful tuning parameter \cite{lonzarich_quantum_2005}.  Unfortunately, for many materials pressures of well above 20 kbar (2 GPa) are necessary to influence the electronic behavior substantially, and anvil cells have to be used that pressurize a rather small volume enclosed between two anvils and the gasket \cite{jayaraman_ultrahigh_1986}. Consequently, sensitivity and accessibility are often an issue for various methods, among them Nuclear Magnetic Resonance (NMR) \cite{harris_high-pressure_2008}. This is the primary reason why only very few NMR studies were carried out at pressures beyond those achievable with clamp cell devices (about 35 kbar). On the other hand, it would be desirable to use NMR methods at higher pressures as they allow one to monitor the electronic behavior of the bulk material locally as a function of temperature. 

This is true in particular for the cuprates \cite{schilling_handbook_2007} where NMR showed the existence of a spin pseudogap, early on \cite{alloul_evidence_1989}. For example, in YBa$_2$Cu$_3$O$_{6+x}$ the $^{89}$Y NMR spin shift is $T$-independent (Pauli-like spin susceptibility) for high doping levels ($x \approx 1$), but begins to decrease at increasingly higher $T > T_{\rm c}$ as $x$ is reduced, despite the fact that the superconducting transition temperature $T_{\rm c}$ decreases. 

With the pseudo-gap phenomenon still unresolved, clearly, it would be advantageous to also investigate the NMR spin shift as a function of pressure and not just of doping. Indeed, there have been attempts but the available pressures were not high enough to observe substantial changes in the NMR parameters \cite{zheng_63Cu_1995, machi_nqr_2003}, so far.

In order to remedy this situation, we have introduced a new anvil cell design for NMR, recently \cite{haase_high_2009}, and showed that this is indeed a promising approach for pressures up to at least 100 kbar \cite{meissner_new_2010, goh_aluminum_2011}. Here we report on the first application of the method to the investigation of the spin shift pseudo-gap of the $^{17}$O NMR of stoichiometric YBa$_2$Cu$_4$O$_8$ \cite{mangelschots_17o_1992} at pressures up to 63 kbar. The $^{17}$O NMR spin shift is particularly easy to interpret as it measures directly the uniform spin susceptibility, since orbital and quadrupolar shifts are vanishingly small and shift measurements do not require assumptions about particular hyperfine scenarios.

The YBa$_2$Cu$_4$O$_8$ powder sample was prepared as described in \cite{williams_nmr_2000}. Measurements of the d.c. magnetization in a field of 2~mT yielded a superconducting transition temperature at ambient pressure of $T_{\rm c} = 81\ \mbox{K}$.  $^{17}$O exchange was performed on the powder pellet at $700\ ^{\circ}\mbox{C}$ in $70 \% $ enriched O$_2$ gas for several hours. It was not possible to align the powder for high pressure experiments, therefore all experiments were performed on pellet chips of the unaligned powder. 

\begin{figure}
\includegraphics[width=0.3\textwidth]{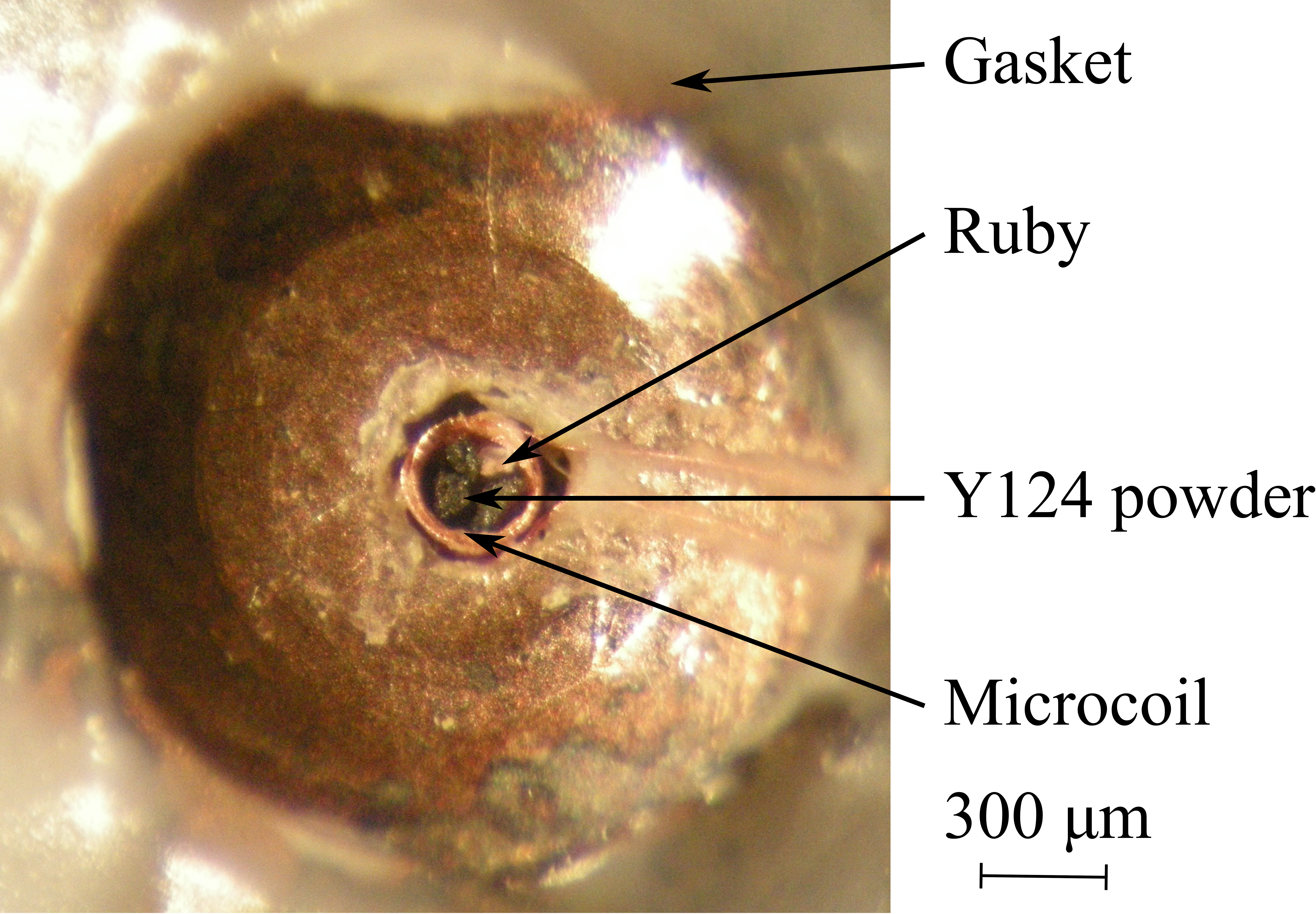}
\caption{(Color online) Photograph, top view along the  axis of the cylindrical NMR radio frequency coil of 300 $\mu \mbox{m}$ diameter (in the middle of the picture) on the gasket of an opened moissanite anvil cell. The coil contains YBa$_2$Cu$_4$O$_8$ (Y124) powder (pieces from a pellet) and a small ruby chip for pressure measurements.}
\label{fig:details}
\end{figure}

Two Dunstan type moissanite anvil cells (MACs) made of hardened non-magnetic beryllium copper (BeCu) were used \cite{dunstan_technology_1989} and pressurized initially to 20~kbar and 42~kbar. The moissanite anvils had culet diameters of 1.0~mm (20~kbar) and 0.8~mm (42~kbar). The gasket was made of ultra pure BeCu with initial thickness of $550\ \mu\mbox{m}$. The gasket was pre-indented to $160\ \mu\mbox{m}$ and a hole of $400\ \mu \mbox{m}$ was drilled at the center of the gasket to accommodate the 10 turn micro-coil wound of $12\ \mu\mbox{m}$ diameter Cu wire. Pieces of the powder pellet and small ruby chips were placed inside the micro-coil. The gasket hole was flooded with glycerin to ensure almost hydrostatic conditions \cite{osakabe_feasibility_2008}. A photograph showing the center part of the gasket before closing of the MAC is displayed in Fig.~\ref{fig:details}. Pressure was measured via the ruby fluorescence method \cite{jayaraman_ultrahigh_1986}.

A particular MAC was then mounted on a home-built NMR probe that fits regular cryostats for $T$-dependent NMR measurements with standard wide-bore superconducting magnets. Further details of the setup are described elsewhere \cite{haase_high_2009}. After the NMR measurements with the 20~kbar and 42~kbar cells the pressure of the latter was increased to 63~kbar.

The NMR shift measurements reported here were performed in a magnetic field of $B_0 = 11.74\ \mbox{T}$ in the temperature range between 300 $\mbox{K}$ and about $85\ \mbox{K}$. The spectra were obtained using the spin-echo sequence  $t_{\pi/2} - \tau - t_{\pi}$ with $t_{\pi/2} = 1.7\ \mu\mbox{s}$ and $\tau = 30\ \mu \mbox{s}$. Shifts are referenced to the resonance frequency $ \nu_{\rm ref}$ of $^{17}$O in tap water. Typical number of scans at 20~kbar and 140~K were 400000 with a last delay of 130~ms.

The actual superconducting transition temperature $T_{\rm c}$ of the sample in the pressurized cell was measured in zero field by monitoring the change in effective inductance of the NMR coil via the resonance frequency of the NMR circuit. We found $T_{\rm c}$'s of 92, 102 and 103~K at 20, 42 and 63~kbar, respectively. The values are in agreement with what has been reported in the literature  \cite{eenige_superconductivity_1990}. 

\begin{figure}
\includegraphics[width=0.384\textwidth]{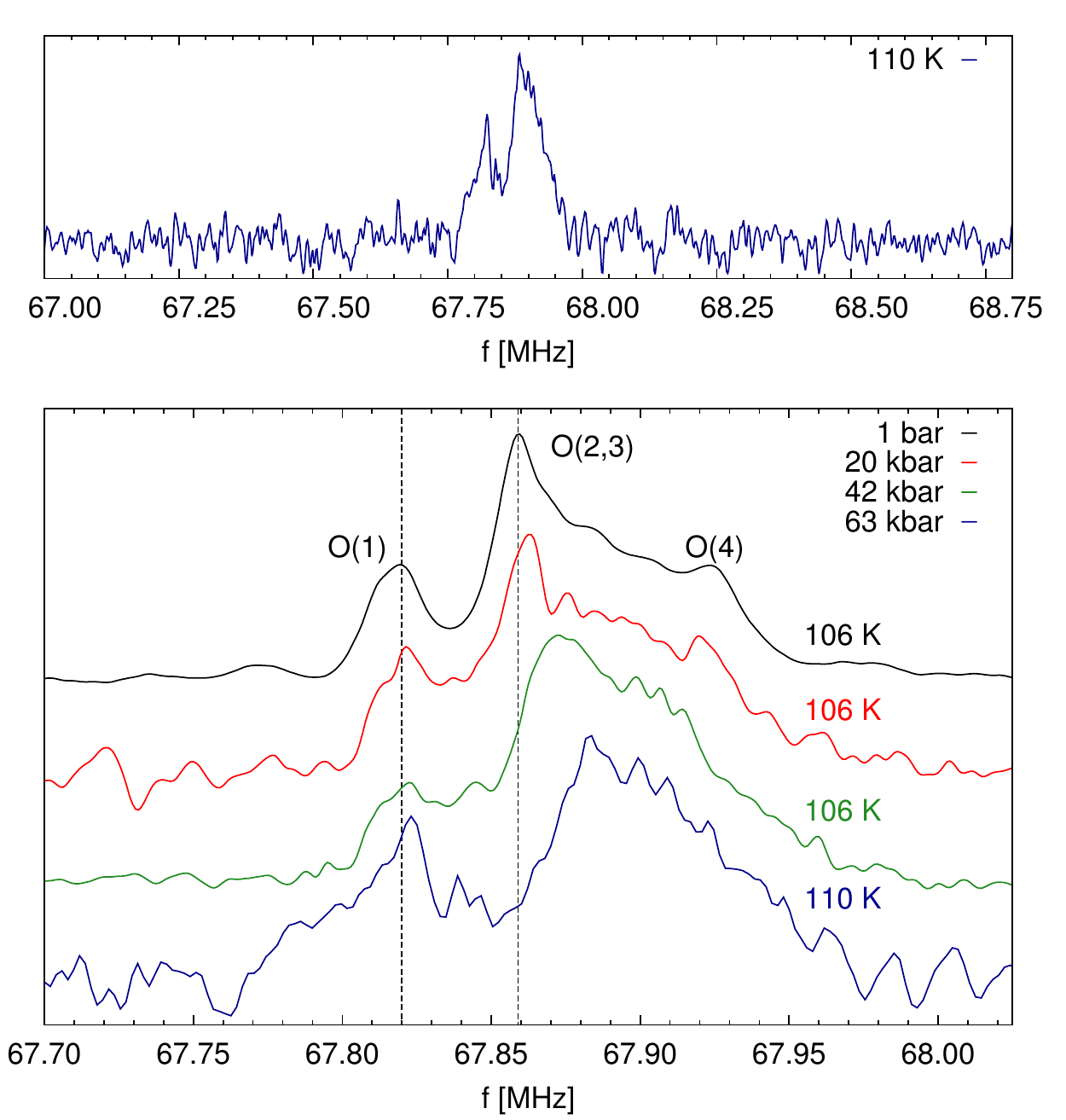}
\caption{\label{fig:Examples} (Color online) Selected $^{17}$O NMR spectra. Top panel, broad band spectrum at 63~kbar, and, bottom panel, narrow band spectra at ambient pressure (1~bar) and high pressures (20, 42 and 63~kbar), in a magnetic field of $B_0 = 11.74\ \mbox{T}$ at the temperatures given in the figure. The peaks observed at ambient pressure are assigned to the apex oxygen O(1), plane oxygen O(2,3) and chain oxygen site O(4). No significant changes in resonance frequency are observed up to 63~kbar for O(1) while the resonance frequency for O(2,3) increases significantly with higher pressure. Dashed lines are a guide to the eye.}
\end{figure}

Typical $^{17}$O NMR powder spectra recorded with the anvil cells are shown in Fig.~\ref{fig:Examples}. Three distinct peaks are visible at ambient pressure that are readily assigned to the apex oxygen O(1), plane oxygen O(2,3), and the chain oxygen O(4) sites \cite{zheng_17o_1992}.  Note that at the given resolution in Fig.~\ref{fig:Examples} we cannot distinguish between the two planar oxygen resonances O(2,3). While the apical O(1) and planar oxygen O(2,3) sites can easily be identified at all pressures, at 42 and 63~kbar we could not clearly resolve the O(4) signal. We notice strong changes in the resonance frequency of the O(2,3) site, while that of the O(1) is hardly affected by increasing the pressure.

The resonance frequency $ \nu$ of a particular $^{17}$O site is influenced by electronic orbital and spin effects, as well as the electric quadrupole interaction of the $I=5/2$ nucleus with quadrupole moment $e$Q situated in a local electric field gradient with its largest principle axis value $V_{\tiny \mbox{ZZ}}$. The resulting quadrupole frequency $2 \pi \nu_{\tiny \mbox{Q}}=\omega _{\tiny \mbox{Q}}  = {{3e{\mbox{Q}}V_{\tiny \mbox{ZZ}} }} / {{2I\left( {2I - 1} \right)}}$ \cite{slichter_principles_1990} for the various oxygen sites can be found in the literature \cite{zheng_17o_1992}. For the planar oxygen $\nu_Q \approx 730\ \mbox{kHz}$ and it splits the nuclear $^{17}$O resonance in our high magnetic field into $2I=5$ lines. Only the central line is observed for powders as its position is affected by 2nd order effects only (less than about 11 kHz linewidth at $B_{\rm 0} $= 11.74~T and center of gravity shift by $\Delta \nu = {{\nu _Q^2 }}  \left[ {I\left( {I + 1} \right) - 3/4} \right]\left( {1 + \eta ^2 /3} \right) / 30 {\nu _0 }  \approx 2.1\   \mbox{kHz}~\approx~0.003 \%$ at 11.74 T). The  electric field gradient tensors for O(2) and O(3) have their largest principal axis value along the Cu-O-Cu bond and an asymmetry of about $\eta_{\rm q} = 0.213$ and 0.228 \cite{mangelschots_17o_1992}. Thus, slight changes in the quadrupole interaction under pressure due to a reduction of the lattice constants \cite{calamiotou_pressure-induced_2009} (no structural phase transitions were observed in the pressure range investigated) or even stronger changes due to variation of the hole distribution \cite{haase_planar_2004} cannot explain the change in the resonance frequency that we observe in Fig.~\ref{fig:Examples}.

The magnetic shifts $K= (\nu - \nu_{\rm ref})/ \nu_{\rm ref}$ for the oxygen central transitions in the normal state have an orbital and spin component, i.e., $K=K_{\rm L} + K_{\rm S}$. The orbital term $K_{\rm L}$ at ambient pressure for O(2,3) is rather small, $K_{\rm L} \approx 0.007\%$ \cite{takigawa_cu_1991, mangelschots_17o_1992}. Therefore, our observed shift changes due to pressure ($p$) or temperature ($T$) must be changes of the spin shift $K_{\rm S}(p,T)$ and thus the spin susceptibility $\chi(p,T)$. The spin shift tensor components for O(2,3) have been determined earlier; the largest is along the Cu-O-Cu bond axis and typical values at 100~K are $K_{\rm iso}=0.10\%, \delta = 0.056\%, \eta = 0.18$ \cite{mangelschots_17o_1992}, for the isotropic shift, anisotropic shift, and asymmetry of the shift tensor, respectively.

\begin{figure}
\includegraphics[width=0.48\textwidth]{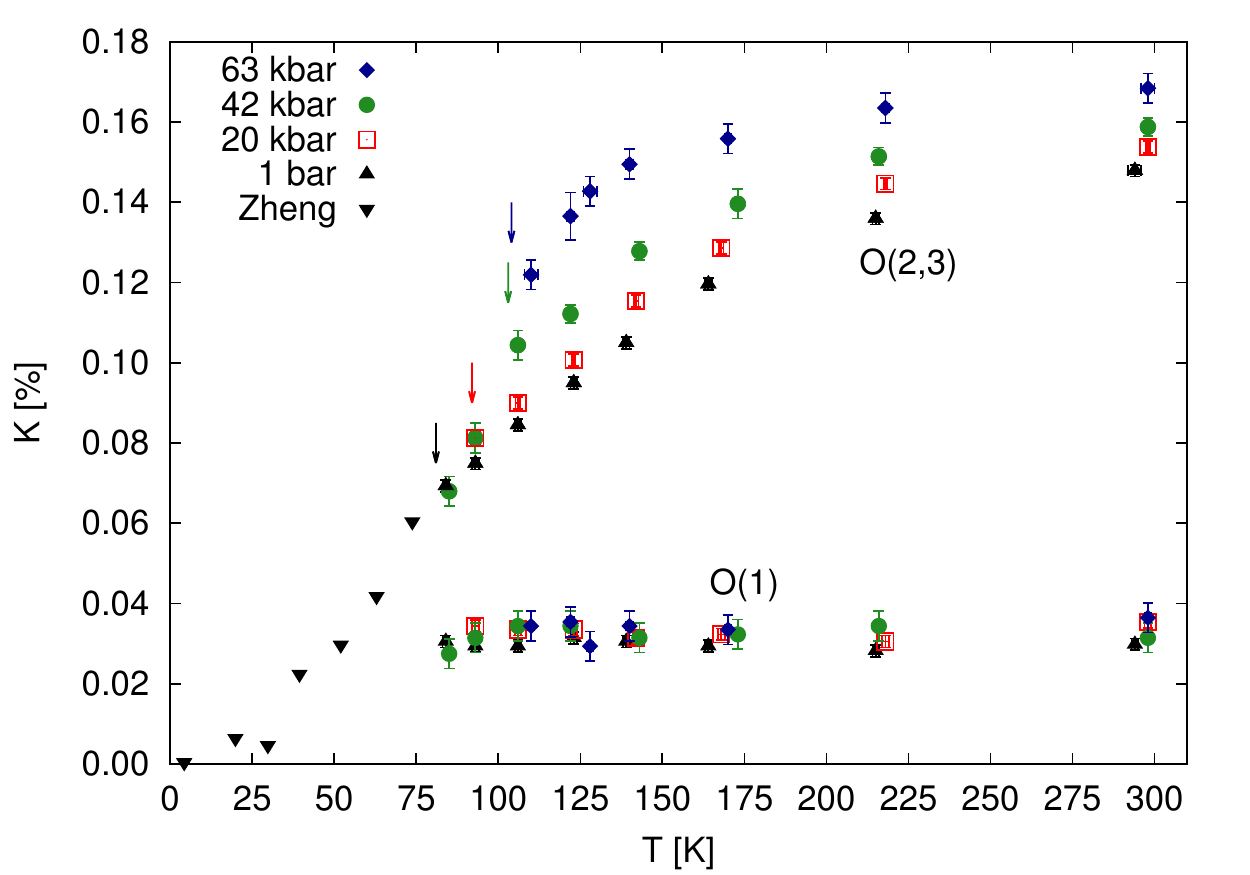}
\caption{\label{fig:Shifts} (Color online) $^{17}$O NMR shifts $K$ as a function of temperature at various pressures. Note that the small, nearly $T$-independent shift values correspond to the apical O(1) nucleus. O(2,3) denotes shifts measured for the planar oxygen. In the present study only data above $T_{\rm c}$ were recorded due to low signal-to-noise; the ambient pressure data below $T_{\rm c}$ (Zheng) are from the literature  \cite{zheng_17o_1992}. The arrows indicate the measured $T_{\rm c}$. Note that pressure increases the shift at all $T$, in particular at lower $T$ such that the pseudo-gap feature begins to vanish with increasing pressure.}
\end{figure}

In Fig.~\ref{fig:Shifts} the magnetic shifts $K(p,T)$ for O(1) and O(2,3) at ambient pressure and 20, 42 and 63~kbar are shown as a function of temperature. The ambient pressure data are in agreement with the literature \cite{zheng_17o_1992, mangelschots_17o_1992}. Due to limitations in signal-to-noise below $T_{\rm c}$ our reported shifts were measured mostly above $T_{\rm c}$. For the ambient pressure shifts we added data below $T_{\rm c}$ from the literature \cite{zheng_17o_1992} in Fig.~\ref{fig:Shifts}. Note that $T_{\rm c}$ is known to increase with pressure and our measured values are indicated by the arrows in Fig.~\ref{fig:Shifts}. We would like to point out that in the normal state no significant change in the signal intensity was observed, ruling out spectral changes as the cause for the measured shift variation (e.g., wipe-out of parts of the signal). The apical oxygen O(1) is only weakly coupled to the electronic fluid in the plane (small hyperfine constant) so that its shift changes little with temperature and pressure compared to O(2,3). 

If coupled to a Fermi liquid the O(2,3) nuclei's spin shift would be $T$-independent down to $T_{ \rm c}$ contrary to what is observed for the ambient pressure shift that decreases already at room temperature. This is the manifestation of the pseudogap in this hallmark pseudogap material. We find that as the pressure increases the shift approaches that of a Fermi liquid, thus the pseudogap gradually disappears in such a way that increasing pressure increases the spin susceptibility at any given temperature in the normal state. Note that some of the low-$T$ points show the influence of $T_{ \rm c}$ as they drop precipitously (see below).

\begin{figure}
\includegraphics[width=0.48\textwidth]{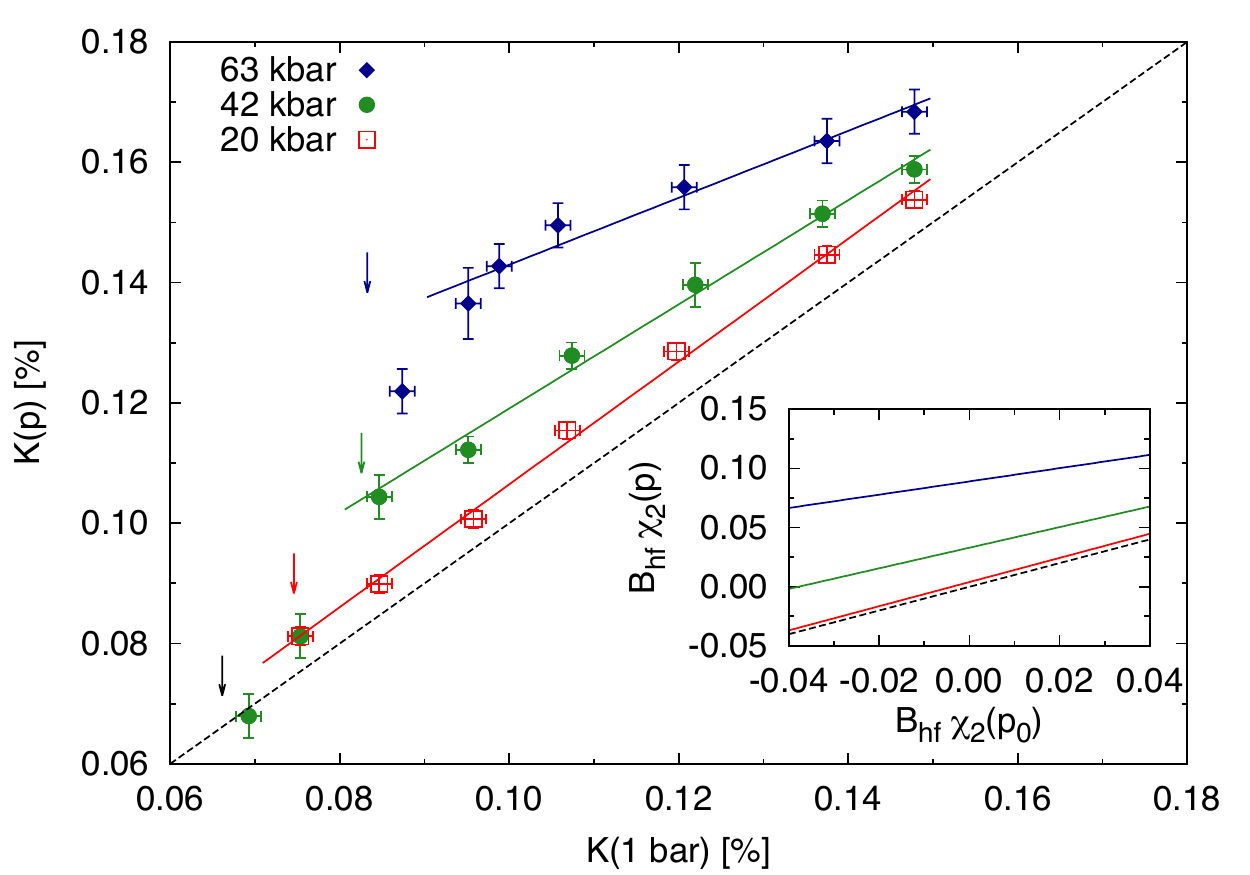}
\caption{\label{fig:DiscussionShifts} (Color online) Main panel: $K(p)$ versus $K(1\ \mbox{bar})$ for O(2,3). Temperature is an implicit parameter. $K(p)$ is linear in $K(1\ \mbox{bar})$ above $T_{\rm c}$ for the investigated pressures. The dashed diagonal has slope 1 for comparison. Note that the lowest shift at 42~kbar is the same as for 1~bar (in this plot the point in the lower left corner), and the 2nd lowest shift at 42~kbar is the same as for 20~kbar (two points overlap near 0.08 and 0.08\%). The arrows indicate the measured $T_{\rm c}$. Inset: ${\rm B}_{\rm hf}\chi_{\rm 2}(p_j)$ vs. ${\rm B}_{\rm hf}\chi_{\rm 2}(p_0).$}
\end{figure}

Inspired by earlier findings on a different system \cite{rybicki_spatial_2008} we plot in Fig.~\ref{fig:DiscussionShifts} $K(p_j,T)$ vs. $K(p_0 = 1 {\rm bar},T)$  with $T$ as an implicit parameter. Interestingly, we find a linear behavior $K(p_j,T)=\kappa_{j,0} \cdot K(p_0,T) + c_{j,0}$ for the data above $T_{\rm c}$, i.e., the slopes $\kappa_{j,0}$ and the constants $c_{j,0}$ do not depend on temperature, only on pressure for basically all data points. We determine $\kappa_{j,0} = 1.02,\ 0.87,\ 0.56$, and $c_{j,0} = 0.004,\ 0.033,\ 0.089 \%$ for $p_j = 20,\ 42,\ 63$~kbar, respectively. The linearity says that the ratio $ \kappa_{j,k}\equiv \Delta K(p_j) / \Delta K(p_k)$ is independent of $T_a, T_b > T_{\rm c}$, where $\Delta K(p_j) \equiv K(p_j,T_b) - K(p_j,T_a)$. This is remarkable since $\Delta K(p_n)$ varies strongly as one changes $T_a$ or $T_b$, cf. Fig.~\ref{fig:Shifts}. The pressure dependent, but $T$-independent constants $c_{j,0}$ demand a $T$-constant (above $T_{\rm c}$) spin shift. 

Note that the spin shift and hence $K(p_j,T)$ in Fig.~\ref{fig:DiscussionShifts} have to disappear below $T_{\rm c}$ so that the high-pressure points in Fig.~\ref{fig:DiscussionShifts} are expected to approach the diagonal (dashed line) below $T_{\rm c}$, eventually. We clearly observe the onset of this behavior for some of the 42~kbar data points (green dots) and at least one point at 63~kbar (blue). The latter point appears to have dropped already above $T_{\rm c}$ (blue arrow), however inhomogeneity of the pressure across the sample could play a role as well as superconducting fluctuations.

Thus, one is lead to a spin susceptibility that is a sum of two terms, one, $\chi_{\rm 1}(p_j,T)$, that is $T$-dependent and decreases proportionally with pressure, and a $T$-constant term, $\chi_{\rm 2}(p_j)$, that depends on pressure only, above $T_{\rm c}$. We thus write
\begin{equation}
\label{eq:TwoComp}
K_{\rm S}(p_j,T>T_{\rm c})={\rm A}_{\rm hf} \chi_{\rm 1}(p_j,T)+{\rm B}_{\rm hf}\chi_{\rm 2}(p_j),
\end{equation}
where ${\rm A}_{\rm hf}, {\rm B}_{\rm hf}$ are the hyperfine (hf) coupling coefficients of the $^{17}$O nucleus to the two spin components.
Note that below $T_{\rm c}$ the spin susceptibility must vanish, or at least become very small, since we must have $K(p_j,T \rightarrow 0) \approx 0$. Although we could not follow this behavior properly by shift measurements below $T_{\rm c}$ due to signal-to-noise limitations, we clearly observe in Fig.~\ref{fig:DiscussionShifts} for the higher pressures a pronounced drop of data points in the vicinity of $T_{\rm c}$. Note that this drop of the high pressure shifts near $T_{\rm c}$, easily recognizable in the main panel of Fig.~\ref{fig:DiscussionShifts}, most likely represent a rapid change of the 2nd component, only, given the scaling law and the almost smooth change of $\chi_1(p_0,T)$ through $T_{\rm c}$, cf. Fig.~\ref{fig:Shifts}. Then, the drop of the 42 kbar data of almost 0.04\% tells us that ${\rm B}_{\rm hf}\Delta\chi_{\rm 2}$(42 kbar)$\approx 0.04\%$. From the determined constants $\kappa_{j,0}$ and $c_{j,0}$ we can now estimate ${\rm B}_{\rm hf}\Delta\chi_{\rm 2}$ for the other pressures. In the inset of Fig.~\ref{fig:DiscussionShifts} we plot ${\rm B}_{\rm hf}\chi_{\rm 2}(p_j)$ as a function of ${\rm B}_{\rm hf}\chi_{\rm 2}(p_0)$ in the range of $-$0.04\% to $+$0.04\%, from which we conclude that ${\rm B}_{\rm hf}\chi_{\rm 2}(p_j)$ is about +0.01\%, +0.01\%, +0.04\%, +0.09\% for pressures of 1~bar, 20~kbar, 42~kbar, and 63~kbar, respectively. These are substantial changes with pressure.

Note that we were forced to introduce the description in terms of (\ref{eq:TwoComp}) entirely based on the pressure dependence of the planar oxygen shift data above $T_{\rm c}$, which are hardly influenced by neither orbital nor quadrupolar effects and do not suffer from Meissner diamagnetism. In fact, an indication for the failure to explain $^{17}$O shift data on YBa$_2$Cu$_4$O$_8$ by a single susceptibility was put forward based on the shift anisotropy earlier \cite{machi_17O_2000}. 

Surprisingly, our findings here are in agreement with those of Haase, Slichter, and Williams \cite{haase_evidence_2009} who investigated the $^{63}$Cu and $^{17}$O NMR of La$_{1.85}$Sr$_{0.15}$CuO$_4$, and more recently, Rybicki et al.  \cite{rybicki_et_al._be_????} who measured the $^{199}$Hg NMR of  HgBa$_2$CuO$_{4 + \delta }$. Both studies suggest that a two-component spin susceptibility is necessary to explain the data. In addition, they \cite{haase_evidence_2009,rybicki_et_al._be_????} also find that one component carries the $T$-dependence of the spin shift pseudo-gap and thus is $T$-dependent already at much higher $T$, but does not change abruptly near $T_{\rm c}$. Their second component is  independent on temperature above $T_{\rm c}$ and disappears rapidly near $T_{\rm c}$, similar to our second component. (In the notation of \cite{haase_evidence_2009} our first component ${\rm A}_{\rm hf} \chi_{\rm 1}(p_j,T)={\rm p}\cdot \chi_{\rm AA}$, and our second component ${\rm B}_{\rm hf}\chi_{\rm 2}={\rm (p+q)}\cdot \chi_{\rm AB}+{\rm q}\cdot \chi_{\rm BB}$, where p, q are the hyperfine coefficients of the nucleus to the two spin components with susceptibilities $\chi_{\rm AA}+\chi_{\rm AB}$ and $\chi_{\rm AB}+\chi_{\rm BB}$ ; for the total uniform susceptibility they have $\chi_{\rm AA}+2\cdot \chi_{\rm AB} +\chi_{\rm BB}$). 

The effect of pressure is notably to increase $T_{\rm c}$ and to trend to a larger and more Pauli-like (i.e. $T$-independent) shift. Both effects taken together seem to be consistent with a rapid suppression of the pseudogap with pressure. From that perspective, the temperature-dependent component we identify from our analysis corresponds to a reconstruction which partially gaps the Fermi surface. However, there appears to be a difference between applying pressure and increasing doping. While hole doping an underdoped cuprate initially \emph{increases} the $T$-dependent component and only slightly the $T$-constant part \cite{alloul_evidence_1989, bobroff_^17o_1997}, applying pressure \emph{reduces} the $T$-dependent component, cf. Fig.~4, but drastically increases the $T$-constant component.

\begin{acknowledgments}
%acknowledgement
\textbf{Acknowledgement}
We thank D. Rybicki, M. Jurkutat, L. Klintberg, M. Lux, S. Sambale,  J. Kohlrautz for help in performing the experiments and J. Tallon and P. Alireza for advice. Financial support by the DFG (International Research Training Group Diffusion in Porous Materials) (TM), Trinity College (Cambridge) (SKG), Deutscher Akademischer Austausch Dienst (PPP USA) and the Royal Society is gratefully acknowledged. 
\end{acknowledgments}

\end{document}